\documentclass[letterpaper,english,reprint,aps,superscriptaddress]{revtex4-1}
\usepackage[T1]{fontenc}
\usepackage[latin9]{inputenc}
\usepackage{amsmath}
\usepackage{amssymb}
\usepackage{graphicx}
\usepackage{soul}
\usepackage{float}
\usepackage{colortbl}
\usepackage[colorlinks=true,citecolor=blue,linkcolor=blue]{hyperref}

\usepackage[lined,boxed,commentsnumbered]{algorithm2e}
\usepackage{tikz}
\SetAlgoCaptionSeparator{.}
\SetAlgoCaptionLayout{justify}

\makeatletter

\pdfpageheight\paperheight
\pdfpagewidth\paperwidth

\usepackage{mathdots}

\newcommand{\ket}[1]{| #1 \rangle}

\newcommand{\expec}[1]{\langle #1 \rangle}

\input{Qcircuit}

\begin{document}

\title{Variational quantum generators: Generative adversarial quantum machine learning for continuous distributions.}

\author{Jonathan Romero}
\email[E-mail:]{jromerofontalvo@g.harvard.edu}
\affiliation{Department of Chemistry and Chemical Biology, Harvard University, 12 Oxford Street, \protect\\ Cambridge, MA 02138, USA}
\affiliation{Zapata Computing Inc., 501 Massachusetts Avenue, Cambridge, MA, 02139, USA.}
\author{Al\'{a}n Aspuru-Guzik}
\email[E-mail:]{aspuru@utoronto.ca}
\affiliation{Zapata Computing Inc., 501 Massachusetts Avenue, Cambridge, MA, 02139, USA.}
\affiliation{Department of Chemistry and Department of Computer Science, University of Toronto, 80 St. George Street, Toronto, Ontario M5S 3H6, Canada.}
\affiliation{Vector Institute, 661 University Ave., Suite 710 Toronto, ON M5G 1M1, Canada.}
\affiliation{Canadian Institute for Advanced Research (CIFAR) Senior Fellow, 661 University Ave., Suite 505, Toronto, ON M5G 1M1, Canada.}

\begin{abstract}
We propose a hybrid quantum-classical approach to model continuous classical probability distributions using a variational quantum circuit. The architecture of the variational circuit consists of two parts: a quantum circuit employed to encode a classical random variable into a quantum state, called the \emph{quantum encoder}, and a variational circuit whose parameters are optimized to mimic a target probability distribution. Samples are generated by measuring the expectation values of a set of operators chosen at the beginning of the calculation. Our \emph{quantum generator} can be complemented with a classical function, such as a neural network, as part of the classical post-processing. We demonstrate the application of the quantum variational generator using a generative adversarial learning approach, where the quantum generator is trained via its interaction with a discriminator model that compares the generated samples with those coming from the real data distribution. We show that our quantum generator is able to learn target probability distributions using either a classical neural network or a variational quantum circuit as the discriminator. Our implementation takes advantage of automatic differentiation tools to perform the optimization of the variational circuits employed. The framework presented here for the design and implementation of variational quantum generators can serve as a blueprint for designing hybrid quantum-classical architectures for other machine learning tasks on near-term quantum devices.
\end{abstract}

\maketitle

\section{Introduction}

Quantum computing, a technology that relies on the properties of quantum systems to process information, is rapidly reaching maturity. Important problems that are hard to solve on classical computers based on transistors, such as factoring and simulating quantum systems, can be solved more efficiently using quantum computers \cite{shor1994algorithms,Lloyd1996UniversalSimulators}.
These devices are nearing the noisy intermediate-scale quantum (NISQ) era \cite{Preskill2018QuantumBeyond}, corresponding to machines with 50 to 100 qubits and capable of executing circuits with depths on the order of thousands of elementary two-qubit operations \cite{Preskill2018QuantumBeyond,wendin2017quantum}. While NISQ devices will not be able to implement error-correction, as opposed to fault-tolerant quantum computers (FTQC), they are expected to provide computational advantages over classical supercomputers for certain problems, which includes sampling from hard-to-simulate probability distributions \cite{AaronsonComplexity-TheoreticExperiments,Boixo2016CharacterizingDevices,Preskill2018QuantumBeyond}.

The limitations in the number of qubits and coherence times of NISQ devices have encouraged the adoption of the hybrid quantum-classical (HQC) framework as the \textit{de facto} strategy to design practical algorithms in the near term. The basic idea behind the HQC framework is that a computational problem can be divided into several subtasks, several of which can be executed more efficiently using a quantum computer while the rest can be deployed to a classical computer. A subset of HQC algorithms called adaptive hybrid quantum-classical (AHQC) algorithms, use classical resources to perform optimization of algorithm parameters. In this case, the quantum subtask generally refers to the process of preparing a parameterized quantum state, followed by the measurement of the expectation values of a polynomial number of observables that encode information relevant to the problem-of-interest. The parameterized quantum state is obtained using a \emph{variational quantum circuit}, which consists of a set of tunable quantum gates whose parameters are subject to optimization. Examples of HQC algorithms include the variational quantum eigensolver (VQE) \cite{Peruzzo2014AProcessor.,McClean2016TheAlgorithms}, the quantum approximate optimization algorithm (QAOA) \cite{farhi2014quantum}, the variational quantum error-correction scheme (QVECTOR) \cite{johnson2017qvector}, among others.

The HQC framework has also been adopted as the basis for designing quantum machine learning algorithms for NISQ devices. One of the first algorithms of this type is the quantum autoencoder (QAE) \cite{Romero2016QuantumData,Wan2016QuantumNetworks}, where a variational quantum circuit is optimized to compress a set of quantum states. This is analogous to a classical autoencoder where an artificial neural network (ANN) is trained to compress classical datasets. The connection between neural networks and variational circuits has been further investigated, where it was shown that the HQC framework can approximate nonlinear functions just as classical neural networks can \cite{cao2017quantum,Mitarai2018QuantumLearning}. Furthermore, variational circuits have provided a new strategy for encoding classical information into quantum states, which is a fundamental step for machine learning applications. In contrast with \emph{amplitude encoding}, in which the input vector is normalized and transformed directly into a quantum state, variational circuits can encode classical data by encoding the input vector as the set of variational circuit parameters \cite{Schuld2018QuantumSpaces,Havlicek2018SupervisedSpaces,Mitarai2018QuantumLearning}. 

In recent months, the combination of the strategies described above for encoding classical data and designing HQC algorithms have led to rapid growth  of publications on quantum machine algorithms for performing both discriminative \cite{Schuld2018Circuit-centricClassifiers,Schuld2018QuantumSpaces,Farhi2018ClassificationProcessors,grant2018hierarchical,Havlicek2018SupervisedSpaces,chen2018universal,Mitarai2018QuantumLearning,sim2018framework} and generative tasks \cite{huggins2018towards,Verdon2017ACircuits,perdomo2017opportunities,benedetti2018generative} on classical data using NISQ devices. 
In machine learning, discriminative models are trained to learn the conditional probability distribution of a target variable $y$ with respect to a set of observations $x$, or $p(y|x)$. In contrast, generative models are trained to learn the joint probability distribution $p(x, y)$, or alternatively, the conditional probability of the observed data with respect to the target variable, $p(x|y)$. Most HQC algorithms for discriminative modeling use a variational quantum classifier \cite{Schuld2018Circuit-centricClassifiers,Farhi2018ClassificationProcessors,grant2018hierarchical,Havlicek2018SupervisedSpaces,Mitarai2018QuantumLearning,sim2018framework}, where a variational circuit is optimized to directly model $p(y|x)$ using training data $\{x_i, y_i\}$. 
Another strategy is to use a variational circuit as a quantum feature map for unsupervised classification with a support vector machine \cite{Havlicek2018SupervisedSpaces,Schuld2018QuantumSpaces}. 
Meanwhile, HQC approaches to generative modeling have focused on modeling discrete probability distributions by using a variational circuit as a \emph{Born machine} \cite{benedetti2018generative,zeng2018learning,situ2018adversarial,liu2018differentiable}. 
Born machines generate samples via projective measurement on
the qubits, for example, by measuring the qubits in the computational basis. While this approach can learn probability distributions for small datasets used for benchmarking, such as Bars-and-Stripes \cite{benedetti2018generative,liu2018differentiable}, as well as quantum circuits for preparation of certain quantum states \cite{benedetti2018generative}, the application of this model to general problems in generative modeling may be difficult due to the exponential scaling of the number of measurements required for sampling the distribution \cite{benedetti2018generative}.

So far, HQC approaches for generative modeling of continuous probability distributions have not been developed. Most industrial applications, such as image and sound generation fall into this category. In this paper we present a variational circuit architecture designed to generate continuous probability distributions. This \emph{variational quantum generator} (VQG) comprises two quantum circuit components: the first one consists of a parameterized quantum circuit used to encode a classical random variable to a quantum state. The second circuit corresponds to a variational circuit whose parameters are optimized to mimic the target classical probability distribution. The output distribution is obtained by measuring the expectation values of a set of predefined operators, whose values can be post-processed using a classical function. This construction provides considerable flexibility in the design of the variational circuit, allowing to easily incorporate VQG into classical neural network architectures. Furthermore, we show that our VQG architecture can be trained using an adversarial learning approach \cite{goodfellow2014generative,Lloyd2018QuantumLearning} leveraging automatic differentiation \cite{neidinger2010introduction,baydin2018automatic,bergholm2018pennylane} to perform gradient-based optimization. That is, our VQG architecture learns to generate samples from the data distribution based on feedback obtained from a discriminator model, which simultaneously learns to distinguish between the samples coming from the real data distribution and those produced by the generator. We show that VQG can be trained using a classical neural network as well as a variational quantum classifier as discriminators.

Our paper is organized as follows: Section \ref{sec:background} briefly describes generative learning using generative adversarial networks and summarizes some proposals for generative learning on quantum computers. Section \ref{sec:quantum-generator} describes the VQG architecture, its implementation, cost analysis, and training process using adversarial learning. In Section \ref{sec:numerical-simulations} we provide a proof-of-principle implementation and numerical simulation of a VQG example and describe the main challenges for its implementation on NISQ devices. Section \ref{sec:conclusions} offers some concluding remarks.




\section{Background}\label{sec:background}

\subsection{Classical and quantum generative adversarial learning (GANs)}\label{classicalGAN}

The machine learning literature provides a variety of generative models. Most of them are trained using the principle of maximum likelihood, that consists of taking several samples from the data generating distribution to form a training set and changing the parameters of the model to maximize the likelihood of the observed data of being generated by the model. Generative models in machine learning can be classified as explicit or implicit, depending on whether or not a explicit form of the probability density function is used \cite{goodfellow2016nips}. Very few tractable explicit models are known, and most of them rely on approximations to the density function. On the other hand, most of the implicit models consist of approximations that can mimic the process of sampling from the generating distribution. Implicit models are further classified into models that require several steps to generate a single sample, such as Markov chains, and models that can generate a sample in a single step. Generative adversarial networks (GANs) belong to the latter category. 

\begin{figure}[t]
\begin{center}
\includegraphics[width=8cm]{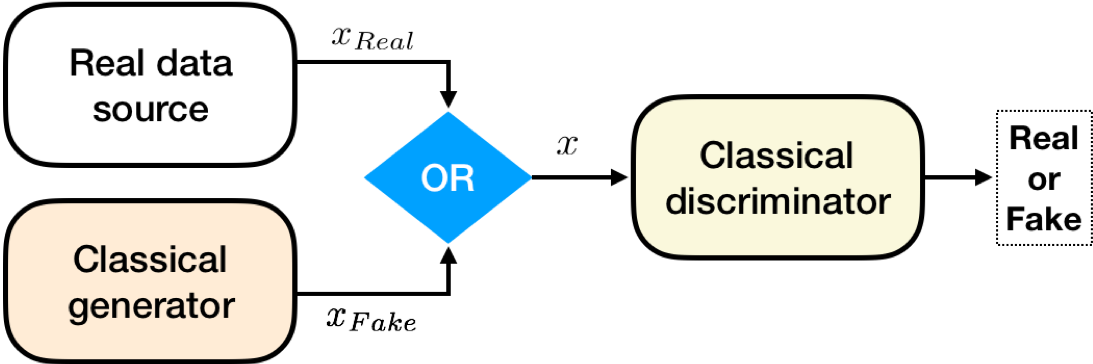}
\end{center}
\caption{Depiction of the classical generative adversarial networks (GANs) scheme: the generator, equipped with random samples from a prior distribution (noise source), produces samples that attempt to mimic the real data samples. The discriminator outputs the probability that a given sample came from the real distribution rather than the synthetic one.}
\label{classicalGANs}
\end{figure}

Generally, GANs consist of two neural networks, the discriminator and the generator, competing against each other in a zero-sum game. Figure \ref{classicalGANs} illustrates the general framework of GANs. Given a prior distribution over the noise parameters $p_z(z)$, the generator consists of a neural network $F_G(z;\Theta_g)$ over the parameters $\Theta_g$ that generates the distribution $p_g$. On the other hand, the discriminator is another neural network $F_D(x;\Theta_d)$ that outputs a single scalar corresponding to the probability of $x$ coming from the real data distribution. Accordingly, $F_D$ is trained to maximize the probability of assigning the correct label to both the training examples and examples coming from $F_G$. Simultaneously, $F_G$ is trained to minimize $log(1-F_D(F_G(z)))$, related to the probability of fooling the discriminator. In summary, $F_D$ and $F_G$ play the following adversarial game:
\begin{align}\label{eq:cf-gan}
&\min_{G} \max_{D}( \mathbb{E}_{x\sim p_{data}(x)}[\log F_D(x)] \notag \\
& + \mathbb{E}_{z\sim p_{z}(z)}[\log(1-F_D(F_G(z)))] ).
\end{align}

Assuming that the discriminator and the generator have infinite capacity, meaning that they can represent any probability distribution, it is possible to show that the final stage of the game reaches a Nash equilibrium where the generator produces data that corresponds to the observed probability distribution, and the discriminator has 1/2 probability of discriminating correctly. Therefore, the final result of the GAN is a generator model that produces samples from the observed distribution by sampling the prior distribution $p_z(z)$. The space of $z$ is usually called the latent space, and $F_G$ is said to map samples from the latent space to the output space $x$. The adversarial framework has proven very successful at training the generator to model a variety of probability distributions, leading to practical applications in many fields, including image synthesis, semantic image editing, molecular discovery, among others \cite{radford2015unsupervised,creswell2018generative,gomez2016automatic}. Nowadays, the application of GANs constitute an exciting and fast growing research field that promises to impact many industries such as self-driving cars, finance, and drug and materials discovery \cite{aspuru2018matter,guimaraes2017objective,sanchez2018inverse,sanchez-lengeling_outeiral_guimaraes_aspuru-guzik_2017,kadurin2017drugan}.

\begin{figure}[t]
\begin{center}
\includegraphics[width=8.5cm]{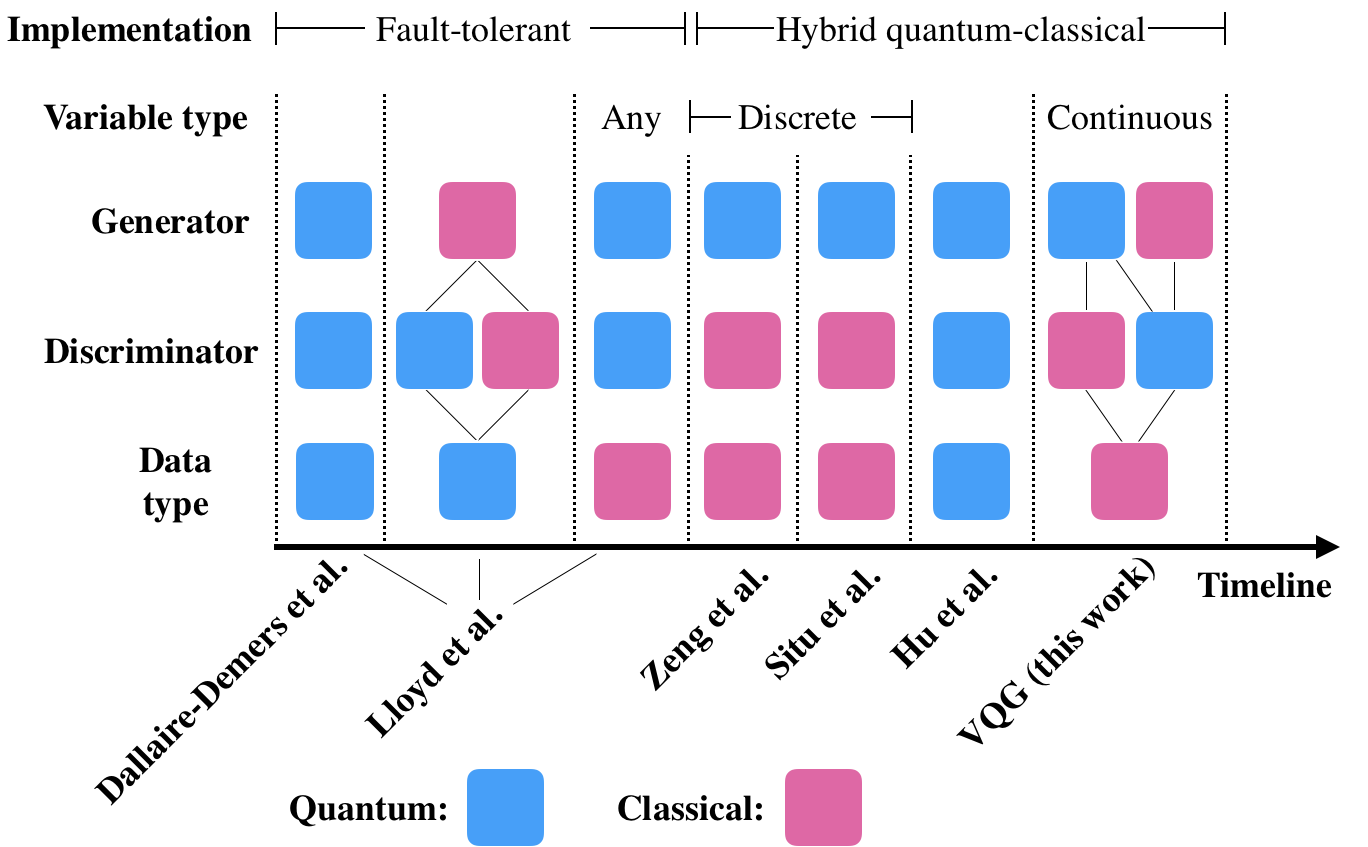}
\end{center}
\caption{Timeline of the development of quantum generative adversarial network models (Dallaire-Demers et al. \cite{Dallaire-Demers2018QuantumNetworks}, Lloyd et al. \cite{Lloyd2018QuantumLearning}, Zeng et al. \cite{zeng2018learning}, Situ et al. \cite{situ2018adversarial}, Hu et al. \cite{hu2018quantum} and this work (VQG)). We describe each proposal in terms of the nature of the data, the discriminator and the generator, that can be either classical or quantum. Lines indicate possible combinations of models and data type. For those models where the type of the data generated is classical, we describe whether the type of variable is discrete or continuous. We also describe the type of implementation proposed, whether it is based on a fault-tolerant model or a hybrid quantum-classical one.}
\label{fig:preliminarwork}
\end{figure}

Recently, different quantum adaptations of the GAN scheme have been proposed \cite{Dallaire-Demers2018QuantumNetworks,Lloyd2018QuantumLearning,zeng2018learning,situ2018adversarial,hu2018quantum}. These methodologies can be characterized according to whether the data source and the models used as discriminator and generator are classical or quantum. The different scenarios considered so far are summarized in Figure \ref{fig:preliminarwork}. In particular, Ref. \cite{Lloyd2018QuantumLearning} offers a theoretical perspective on three possible adversarial learning scenarios. The first of these settings corresponds to a purely quantum version of GANs, where the data distribution is a quantum source and the models correspond to quantum circuits. This proposal is further developed in \cite{Dallaire-Demers2018QuantumNetworks}, and experimentally demonstrated for a proof-of-principle quantum computation with a superconducting qubit architecture \cite{hu2018quantum}. The second scenario considers a classical generator that is trying to produce quantum data at an exponential cost. The third scenario corresponds to classical data encoded in the amplitudes of a quantum state, such as quantum generators and discriminator can be employed. As described in \cite{Lloyd2018QuantumLearning}, these proposals are designed for error-corrected quantum computers. More recently, some groups have proposed hybrid-quantum classical adversarial learning schemes that could be implemented on NISQ devices. These approaches utilize a classical data source and a classical discriminator combined with a variational circuit sampled as a Born machine as generator \cite{zeng2018learning,situ2018adversarial}. As noted earlier, the Born machine approach consists of generating a discrete distribution via projective measurement on the qubits.

\section{The variational quantum generator architecture}\label{sec:quantum-generator}

\subsection{Architecture}

\begin{figure}[t]
\begin{center}
\includegraphics[width=8cm]{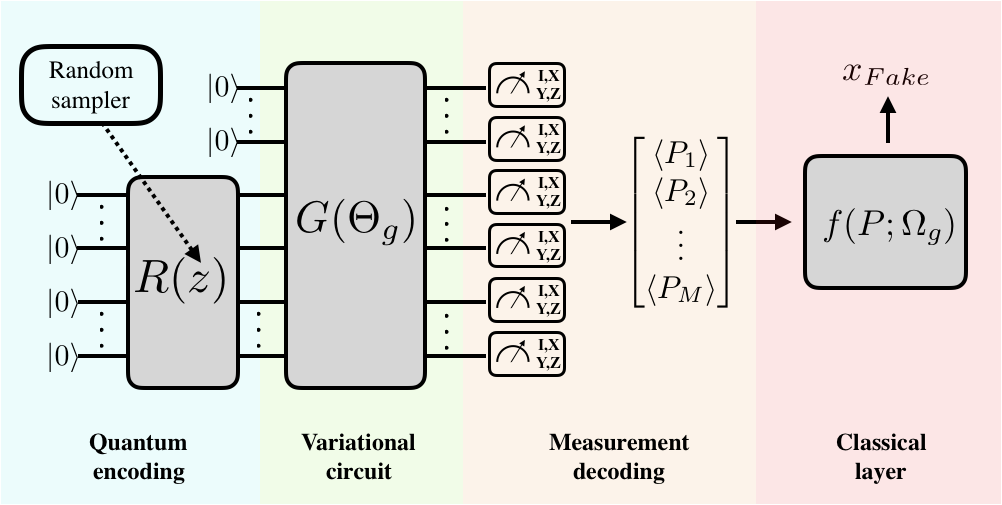}
\end{center}
\caption{Circuit architecture of the proposed quantum generator, comprising a circuit that generates states from a latent space ($z$) using the variational circuit $G(\Theta_g)$. The random variable $z$ is mapped to a quantum state using the quantum encoding circuit $R$. By measuring a fixed set of operators on the generated state, the quantum circuit produces a classical vector $P=[\expec{P_1}, \cdots, \expec{P_M}]$, that passes through a classical function $f(P; \Omega_{g})$, to produce the fake sample $x_{Fake}$.}
\label{VQG}
\end{figure}

Existing quantum models for generative learning collect data by measuring the system as a Born machine, which is convenient for discrete distributions but can cannot be easily adapted for continuous cases. We propose a scheme to generate continuous distributions that builds on the principles of HQC computing. Consider a real data source that outputs observations of an unknown distribution, represented by the variable $x \in \mathbb{R}^{N}$. The purpose of our variational quantum generator is to produce classical samples $x_{Fake}$ that mimic the observed distribution. To achieve this, we propose the construction depicted in Figure \ref{VQG}, that includes two variational circuits, a quantum encoding circuit $R(z)$ acting on $r$ qubits and the generator circuit $G(\Theta_g)$ acting on $n$ qubits with $n \geq r$.

The quantum encoding circuit, which we describe in detail in the next subsection, takes as input a classical random variable $z \sim p_z(z); z \in \mathbb{R}^{O}$ as a parameter and prepares the state $R(z)\ket{0^{\otimes r}}=\ket{\phi(z)}$. This is the equivalent to the random source employed in classical GANs, where the space of the variable $z$ would correspond to the latent space in the language of generative models. Correspondingly, the manifold of states $\{\ket{\phi(z)}\}$ would constitute the quantum latent space. The second circuit, $G(\Theta_g)$ acts as the generator model, mapping from the latent manifold to the manifold of observed data $x$: $G(\Theta_g)\ket{\phi(x)} = \ket{\psi(z, \Theta_g)}$. To map this state to a classical value we employ a \emph{measurement decoding} scheme, where the sample $P \in \mathcal{R}^{M}$ is generated by measuring the expectation value of a fixed set of observables expressed as strings of Pauli strings $\{ P_i\}_{i=1,\cdots,M}$:
\begin{align}
&P = [\expec{P_1}, \expec{P_2}, \cdots, \expec{P_M}] \\
&\text{where} \quad \expec{P_i} = \expec{\psi(x; \Theta_g) |P_i|\psi(x; \Theta_g)}.
\end{align}
$P$ is then transformed by a classical function to generate $x_{Fake}$:
\begin{align}
&x_{Fake} = f_g \left(P; \Omega_g \right), \\
\end{align}
where $\Omega_g$ represents a vector of real parameters associated to the classical function. In what follows, we describe each of the components of VQG in greater detail.

\subsubsection{Quantum encoding circuit}

The process of encoding classical inputs in a quantum state can be interpreted as applying a nonlinear feature map that maps data to a quantum Hilbert space, a process also called \emph{quantum feature map} or \emph{quantum encoding}, as described by Schuld et al. \cite{Schuld2018QuantumSpaces}. The quantum circuit implementing this mapping on a digital quantum computer corresponds to the \emph{quantum feature circuit} or \emph{encoding circuit}. We distinguish between two classes of quantum encoding in this paper:
\begin{enumerate}
    \item \emph{Amplitude encoding}: In the first case, a vector $x \in \mathbb{R}^{N}$, corresponding to the data to be encoded, undergoes a transformation under a feature map: $\psi: \mathbb{R}^{N} \rightarrow \mathbb{C}^{2^n}$ that maps the information to a quantum state in $n$ qubits. Since the length of the data vector is not necessarily a power of 2, the feature map might require some padding and appropriate normalization. Once the corresponding input state is obtained, we need to prepare this state on the quantum register of $n$ qubits, $\ket{\phi(x)}$ using a preparation circuit $S_{x}$ such that $S_{x}\ket{0}^{\otimes n} = \ket{\psi(x)}$. 
    \item \emph{Variational encoding}: In this case, a fixed variational circuit $E(f_E(x))$ encodes the data by inputting the classical information as the circuit parameters. Here, $f$ is a classical feature map: $f_E: \mathbb{R}^{N} \rightarrow \mathbb{R}^{M}$, such as the final input state is prepared as $E(f(x))\ket{0}^{\otimes n} = \ket{\phi(x)}$.
\end{enumerate}

Notice that in amplitude encoding, the vector is mapped classically to a quantum state. Consequently, we need to find the corresponding circuit that prepares the state to a desired accuracy. This can be done using general purpose compilation routines for preparing general quantum states on quantum registers \cite{grover2002creating,soklakov2006efficient,plesch2011quantum,niemann2016}. In the case of amplitude encoding, the number of qubits required scales as $O(\log(N))$ while the depth of the circuit for state preparation is $O(N)$ \cite{plesch2011quantum}, with $N$ being the size of the classical vector to be mapped. The number of gates required for state preparation of these circuits (In the order of thousands for ten qubits \cite{niemann2016}) might constitute a challenge for NISQ devices.

In contrast, the variational encoding strategy encodes the classical vector as the parameters of a fixed variational circuit. This implies that the circuit layout employed for all the input vectors is the same, which simplifies compilation. It is also likely that the errors introduced by this encoding procedure are mostly systematic and therefore can be more easily mitigated. Most of the variational encodings proposed so far employ circuits with $O(N)$ qubits and only $O(1)$ circuit depth \cite{Mitarai2018QuantumLearning,grant2018hierarchical,Schuld2018QuantumSpaces,Schuld2018Circuit-centricClassifiers,huggins2018towards,grant2018hierarchical}, which makes encoding more amenable to NISQ devices at the cost of increasing requirement in the number of qubits compared to amplitude encoding. 

Both amplitude encoding and variational encodings have been used in machine learning proposals for classification \cite{Biamonte2016QuantumLearning,Schuld2018Circuit-centricClassifiers,Havlicek2018SupervisedSpaces,Schuld2018QuantumSpaces,Farhi2018ClassificationProcessors,Mitarai2018QuantumLearning}, and can be employed as part of the VQG architecture. In the space of variational encodings, some specific classes of circuits have been proposed. Some examples include \emph{product encoding}, in which each element of the vector $x$ is mapped to a one qubit state by a specific quantum circuit \cite{Schuld2018QuantumSpaces,Schuld2018Circuit-centricClassifiers,huggins2018towards,grant2018hierarchical}. Other approaches incorporate more layers of single qubit gates whose parameters are given by the elements of the feature vector, followed by circuit blocks made out of fixed entangling operations \cite{Havlicek2018SupervisedSpaces}. A particular strategy that can be used to introduce non-linearity is the so-called \emph{tensorial mapping}, which consists of preparing several copies of the quantum state encoding the data \cite{Mitarai2018QuantumLearning,Schuld2018QuantumSpaces}. 
An example of a variational encoding combining product encoding and tensorial mapping is the following preparation circuit:
\begin{equation}\label{eq:productencoding}
    U(x) = \prod^{N}_{k} \prod^{n_k}_{i} R^i_{Z}(f(x_k)) R^i_{Y}(g(x_k)),
\end{equation}
where each element of the vector $x$ is mapped by a circuit acting on a fixed number of qubits, $n_k$, and $f$ and $g$ correspond to non-linear activation functions. The notation $R_{V}^{I}(\alpha)= e^{-i\frac{\alpha}{2} V_{I}}$ indicates a general rotation under the operator $V$ acting on the set of qubits $I$. Notice that in the map of equation \ref{eq:productencoding}, non-linearities are introduced via the use of non-linear functions as part of the mapping and by application of the tensorial map.

\subsubsection{Variational circuit}\label{subsubsec:variational-circuits}

$G(\Theta_g)$ plays the role of the variational circuit in our VQG architecture. Most variational circuits are designed to prepare strongly entangled quantum states. This follows the general intuition that the variational circuits employed should be able to map the input data into quantum states that are hard to manipulate on classical computers. In addition, variational circuits must be able to spot different types of correlations in the input data, which requires circuits with the ability to explore Hilbert space sufficiently. 
Variational algorithms such as QAE and QVECTOR have been implemented using quantum circuits composed by a fixed networks of a polynomial number of gates, usually restricted to single-qubit and two-qubit operations, with angles that serve as variational parameters. The pattern defining the network of gates can be seen as a \emph{unit-cell} or \emph{circuit block} that can be repeated to increase the flexibility of the model. The term Multilayer Quantum Circuit (MPQC) has been recently coined to describe this type of variational circuit architecture \cite{du2018expressive}. MPQC circuits have been widely used as quantum models for classification tasks \cite{Havlicek2018SupervisedSpaces,Schuld2018Circuit-centricClassifiers} and has been shown to generate discrete probability distributions that cannot be efficiently simulated by classical neural networks \cite{du2018expressive}. We describe the specific architecture of some of these circuits in Appendix \ref{app:VCs}.

Apart from MPQC circuits, it is also possible to use a circuit implementing the evolution under a family of Hamiltonians known to generate strongly correlated states. In this case, the coefficients of the Hamiltonian terms can be used as variational parameters. For instance, Mitarai et al. \cite{Mitarai2018QuantumLearning} used the evolution under a transverse Ising Hamiltonian to perform simulations of quantum classification and to model nonlinear functions using variational circuits \cite{Mitarai2018QuantumLearning}. The circuits implementing evolution under a given Hamiltonian may require Trotterization. In this case, each Trotter step might be interpreted as a circuit block, in analogy with the Hamiltonian variational approach described in Ref. \cite{wecker2015progress}.

\subsubsection{Measurement decoding and post-processing}\label{subsubsec:measurement-decoding}

The process of measurement decoding generates samples from the generator by estimating the vector of expectation values $P=\left[ \expec{P_1}, \expec{P_2}, \cdots, \expec{P_M} \right]$. The choice of operators for decoding depends on the problem at hand and constitutes a hyper-parameter of the model. The cost of estimating the vector $P$ with measurement averaging, assuming each operator is measured independently and with fixed precision $\epsilon$, is $O\left( \sum^{M}_{i=1} \frac{Var[P_i]}{\epsilon^2} \right)$. The associated measurement cost is not different from other HQC algorithms such as VQE, where the expectation value of the Hamiltonian is computed by a weighted average of the expectation value of a polynomial number of observables \cite{McClean2016TheAlgorithms}. Here we assume that $\epsilon$ is independent of subsequent transformations of the vector $P$ and is small enough to carry out the optimization and the generation of samples $x$ successfully.

If the training of the VQG model is carried out with gradient-based optimization, this will require the estimation of $\nabla_{\Theta_G} P$. In this case, the total number of measurements employed depends on the number of circuit runs used per gradient estimation ($N_{grad}$) and the total number of gradient evaluations required by the optimization ($N_{opt}$). As described in Appendix \ref{app:VCs}, $N_{grad} \sim O(N_{p}/\epsilon^2)$ where $N_{p}$ is the total number of parameterized gates in the variational circuit. To generate the final sample $x$, the generator can incorporate a classical function that transforms the measurement vector, $P$. In general, we can express $f_g$ as $f_g(x) = h(WP+b)$, where $W \in \mathbb{R}^{N \times M}$, $b \in \mathbb{R}^{N}$ and $h$ is a function that can be nonlinear $h: \mathbb{R}^{N} \rightarrow \mathbb{R}^{N}$. To unify the notation, we designate $\Omega_{g} = (W, b)$ as in Figure \ref{VQG}. This construction makes VQG a hybrid quantum-classical architecture and therefore the evaluation of the model and its derivatives will require feedback between the classical computer and the quantum processor. We describe this process in more detail in the next section, where we discuss how to train the VQG model.

\subsection{Training and cost function}\label{subsec:costF}

The VQG architecture could be trained by direct maximization of the log-likelihood. However we have chosen to use an adversarial learning approach, which has certain advantages as described in Section \ref{classicalGAN}. The adversarial setting requires a discriminator function $F_D(x; \Theta_d)$, parameterized by $\Theta_d$, that receives the sample $x$ as input and outputs an approximation to the probability of the sample originating from the real distribution. We will describe the architecture of the discriminator shortly, but assuming we can compute the necessary gradients, this discriminator could be trained using the same cost function employed in classical GANs (Eq. \ref{eq:cf-gan}). We rewrite this expression to make explicit the dependency in the parameters:

\begin{align}\label{CFdis}
&C_d(\Theta_d, \Theta_g) = - \frac{1}{2} \mathbb{E}_{x\sim p_{data}(x)}[\log F_D(x; \Theta_d)] \notag \\ & -\frac{1}{2} \mathbb{E}_{z\sim p_{z}(z)}\log[1-F_D( (F_G(z; \Theta_g, \Omega_g); \Theta_d)],
\end{align}
where $F_G(z; \Theta_g, \Omega_g)$ is the function corresponding to VQG. The first term in Equation \ref{CFdis} corresponds to the probability of the discriminator to succeed at classifying data coming from the real source correctly, while the second term represents the probability of the discriminator to succeed at identifying the sample created by the generator as fake. Notice that the discriminator needs to be trained on two batches of data: one corresponding to real samples (for which the discriminator should output 1) and a second batch created by the generator (for which the discriminator should output 0). In classical GANs, the original choice of the cost function of the generator is just the negation of the cost function of the discriminator, such as $C_g(\Theta_d, \Theta_g) = - C_d(\Theta_d, \Theta_g)$, and therefore the final optimization consists of a minimax game:
\begin{align}\label{eq:CF1}
\min_{\Theta_g} \max_{\Theta_d} C_d(\Theta_d, \Theta_g).
\end{align}
At each step of the optimization, the parameters of each player are optimized while the parameters of the other player are kept fixed. One alternative to the cost function in Equation \ref{CFdis} is to use the inverse of the discriminator cost function for the generator, such that $C_g$ becomes:
\begin{equation}\label{eq:CF2}
C_g(\Theta_d, \Theta_g) = -\mathbb{E}_{z\sim p_{z}(z)}[\log F_D( (F_G(z; \Theta_g, \Omega_g); \Theta_d) )].
\end{equation}
In this case, the generator minimizes the probability of the discriminator of being correct. This proposal, although heuristically motivated, has demonstrated the ability to facilitate the training process in classical GANs \cite{goodfellow2016nips}.

\begin{figure}[t]
\begin{center}
\begin{tabular}{c}
Scheme I \\
\includegraphics[width=8cm]{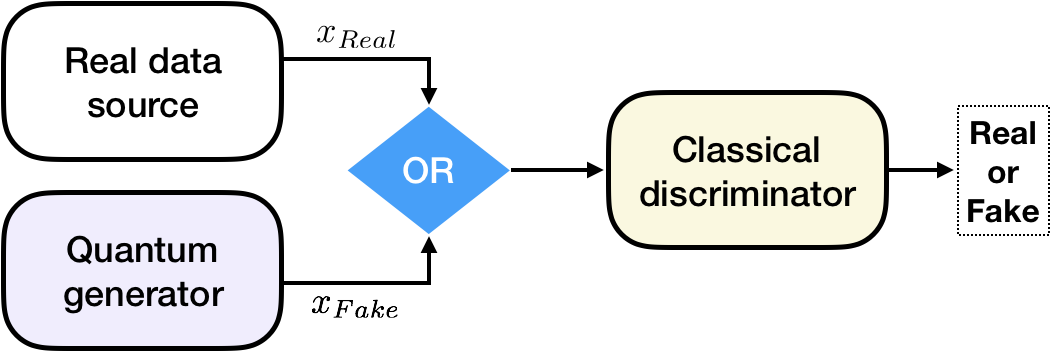} \\
\\
Scheme II\\
\includegraphics[width=8cm]{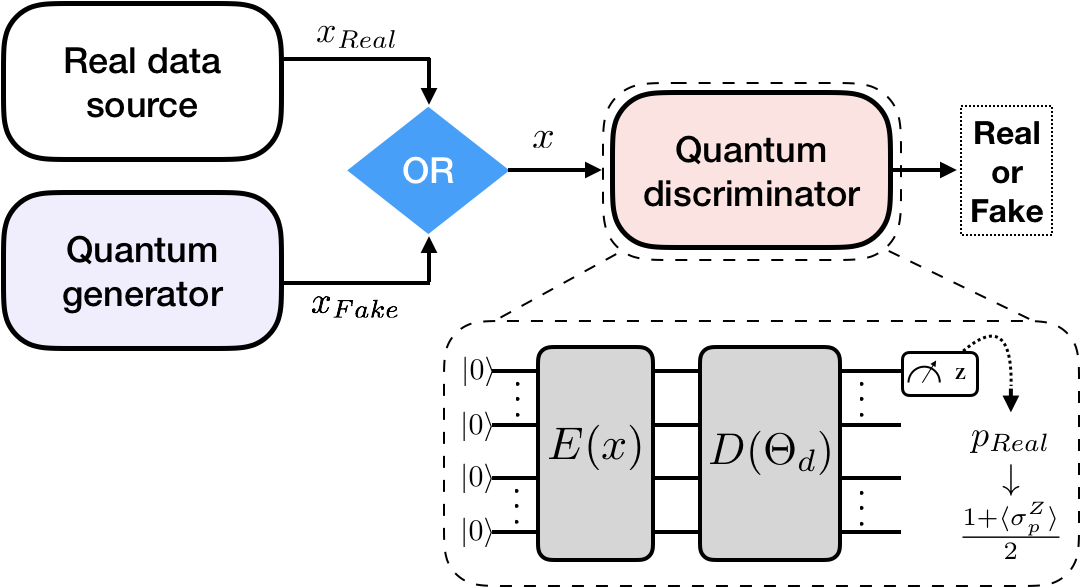}
\end{tabular}
\end{center}
\caption{Two different schemes for training a quantum generator of classical data using an adversarial learning approach: the first scheme (Scheme I) employs a classical discriminator (e.g. classical neural network), whereas the second scheme employs a quantum discriminator (Scheme II), which consists of a quantum circuit that encodes the classical sample ($E(x)$) and a variational circuit ($D(\Theta_d)$), whose parameters are optimized such as the measured observable ($\sigma^{Z}_{p}$) describes the probability of the sample to come from the real distribution.}
\label{fig:HQGAN-schemes}
\end{figure}

As in the case of other quantum machine learning approaches, we propose to use gradient based methods for the optimization. Most recent numerical and experimental demonstrations for classifiers based on variational circuits have employed methods such as simultaneous perturbation stochastic approximation (SPSA) \cite{spall2000adaptive,spall1997one} and stochastic gradient descent (SGD) \cite{goodfellow2016deep}. SPSA is based on numerical gradients and has been already employed in experimental demonstrations of VQE \cite{Kandala2017Hardware-efficientMagnets} and QML algorithms for classification \cite{Havlicek2018SupervisedSpaces}. The difficulty with SPSA is that the number of measurements required increases substantially as the gradient vanishes. In this case, the heuristic cost function for the generator (Equation \ref{eq:CF2}) might require fewer measurements as it prevents vanishing gradients. In contrast, algorithms such as SGD generally work with analytical gradients in the context of classical neural networks. Employing these algorithms for our VQG requires computing gradients with respect to the circuit parameters $\Theta_g$, $\Omega_g$ and $\Theta_d$.

In order to compute gradients for the discriminator we need to define its structure. Since the VQG is designed to generate classical data, it is possible to perform the training using both classical and quantum discriminators. These two possible schemes are pictorially described in Figure \ref{fig:HQGAN-schemes}. In the first scheme, the classification is performed by a classical discriminator, for example, a classical feed-forward neural network. Consequently, the discriminator can be trained by maximizing the cost function (Eq. \ref{CFdis}) using standard back-propagation techniques for feed-forward neural networks. 

In the second scheme, the classification is performed by quantum discriminator model, for example, a variational quantum classifier (VQC) \cite{Mitarai2018QuantumLearning,Schuld2018Circuit-centricClassifiers,Havlicek2018SupervisedSpaces,sim2018framework}. As the input data is classical, the quantum discriminator comprises a quantum encoding circuit, $E(x)$, that maps the data point $x$ to a quantum state, and a variational circuit $D$, with parameters $\Theta_d$. A set of measurements provide the final values indicating the classification. Correspondingly, the structure of this quantum discriminator becomes analogous to the structure of the quantum generator, with the difference that the classical output produced by measurement needs to be transformed such that it corresponds to a probability distribution instead of an arbitrary vector. For the GAN implementation, the discriminator performs only binary classification and therefore $p(y|x)$ can be modeled by measuring a single observable, e. g. $\sigma^{Z}_a$, with $a$ being the index of the designated qubit. Correspondingly, the probability of $x$ coming from the true distribution of the data can be estimated simply as $p_{Real} = \frac{1+\langle \sigma^{Z}_p\rangle_D }{2}$, where:
\begin{align}
\expec{\sigma^{Z}_p}_D = \expec{0|E^{\dagger}(x) D^{\dagger}(\Theta_d) \sigma^{Z}_p D(\Theta_d) E(x)|0}.
\end{align}
Consequently the gradient of the discriminator takes the form:
\begin{align}\label{eq:Qdis-grad}
\frac{\partial F_D(x; \Theta^d_i)}{\partial \theta^d_i} =& \frac{1}{2}\frac{\partial \expec{0|E^{\dagger}(x) D^{\dagger}(\Theta_d) \sigma^{Z}_p D(\Theta_d) E(x)|0}}{\partial \theta^d_i},
\end{align}
Eq. \ref{eq:Qdis-grad} can be evaluated using the standard techniques for computing gradients in variational circuits described in Appendix \ref{app:grad-estimation}. On the other hand, training the generator implies computing gradients of both Eq. \ref{eq:CF1} and Eq. \ref{eq:CF2} with respect to the generator parameters, which ultimately requires the calculation of the following derivatives:

\begin{align}\label{eq:VQG-grads}
&\frac{\partial F_D(F_G(z; \Theta_g, \Omega_g); \Theta_d)}{\partial \theta_g^i} = \sum_{l} \sum_{k} \frac{\partial F_D(x_l)}{\partial x_l} \frac{\partial x_l}{\partial \expec{P_k}} \frac{\partial \expec{P_k}}{\partial \theta^{i}_{g}},
\end{align}
\begin{align}\label{eq:VQG-grads2}
&\frac{\partial F_D(F_G(z; \Theta_g, \Omega_g); \Theta_d)}{\partial \omega_g^i} = \sum_{l} \frac{\partial F_D(x_l)}{\partial x_l} \frac{\partial x_l}{\partial \omega^{i}_{g}}, 
\end{align}
where we have used the following notation:
\begin{align}
\Theta_d &= [\theta_{d}^1, \theta_{d}^2, \cdots, \theta_{d}^{|\Theta_d|}]; \\
\Theta_g &= [\theta_{g}^1, \theta_{g}^2, \cdots, \theta_{g}^{|\Theta_g|}]; \\
\Omega_g &= [\omega_{g}^1, \omega_{g}^2, \cdots, \omega_{g}^{|\Omega_g|}]; \\
x &= [x_1, x_2, \cdots, x_N].
\end{align}
Notice that the partial derivatives appearing in Eq. \ref{eq:VQG-grads} are estimated differently depending on the type of discriminator used. In scheme I, $\frac{\partial F_D(x_l)}{\partial x_l}$, $\frac{\partial x_l}{\partial \omega^{i}_{g}}$ and $\frac{\partial F_D(x_l)}{\partial x_l}$ correspond to derivatives of classical functions and can be computed using established backpropagation techniques. In this case, only $\frac{\partial \expec{P_k}}{\partial \theta^{i}_{g}}$ corresponds to a derivative of a variational circuit. In contrast, $\frac{\partial F_D(x_l)}{\partial x_l}$ is also a derivative of a variational circuit in the case of Scheme II:
\begin{align}\label{eq:Qdis-grad2}
\frac{\partial F_D(x; \Theta^d_i)}{\partial x_i} =& \frac{1}{2}\frac{\partial \expec{0|E^{\dagger}(x) D^{\dagger}(\Theta_d) \sigma^{Z}_p D(\Theta_d) E(x)|0}}{\partial x_i},
\end{align}
which implies taking derivatives of the encoding circuit or the corresponding encoding scheme. In case of variational encodings, the same techniques applied to compute the gradients of circuits $G$ and $D$ can be employed for computing Eq. \ref{eq:Qdis-grad2}. If the encoding involves pre-processing $x$ with a classical function, the calculation of the gradient requires further unfolding as with Eq. \ref{eq:VQG-grads}. 

In summary, to train the VQG model using adversarial learning, we need to compute gradients of variational circuits and apply backpropagation for classical functions. We review the calculation of existing techniques for computing analytical gradients of variational circuit in Appendix \ref{app:grad-estimation}. To compute gradients of classical functions, we exploit state of the art automatic differentiation (AD) techniques \cite{neidinger2010introduction,du2018expressive}. AD is an algorithmic strategy to extend a program that computes numerical values of a function such as it can also compute arbitrary derivatives of the same function, as described in Appendix \ref{app:AD}. This technique is widely used in machine learning to perform automatic calculation of derivatives for gradient-based optimization of models such as neural networks. AD also offers a convenient framework to propagate gradients between classical and quantum functions, as described in Ref. \cite{schuld2018evaluating,bergholm2018pennylane}. 

\begin{figure*}[t]
\begin{center}
\includegraphics[width=13cm]{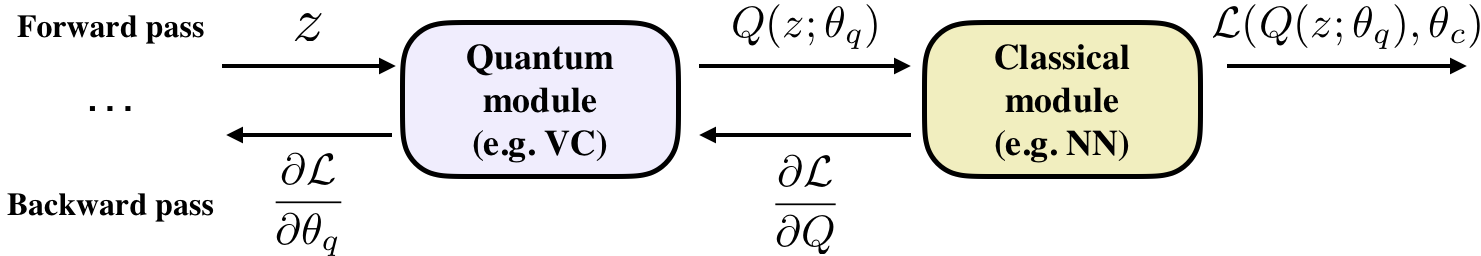}
\end{center}
\caption{\label{fig:QAD} Application of reverse accumulation (See Appendix \ref{app:grad-estimation}) for automatic differentiation of a hybrid quantum-classical architecture. The quantum module implements a function computed from a variational circuit (VC), $Q(x,\theta_q)$, as well as derivatives of this function with respect to $x$ and $\theta_q$. The classical module implements a classical function e.g. a neural network (NN), $\mathcal{L}(y,\theta_c)$, and its derivatives with respect to $y$ and $\theta_c$. In the forward pass, $\mathcal{L}(Q, \theta_c)$ is calculated by first computing $Q(z,\theta_q)$ using the quantum module and passing its value to the classical module. In the backward pass, the classical module computes $\frac{\partial \mathcal{L}}{\partial Q}$ and passes this information to the quantum module. The quantum module estimates $\frac{\partial Q}{\partial \theta_q}$ using the quantum processor and computes $\frac{\partial \mathcal{L}}{\partial \theta_q}$ by application of the chain rule: $\frac{\partial \mathcal{L}}{\partial \theta_q} = \frac{\partial \mathcal{L}}{\partial Q} \frac{\partial Q}{\partial \theta_q}$.}
\end{figure*}

Figure \ref{fig:QAD} illustrates the calculation of gradients for a hybrid-quantum classical function, such as a VQG module, using AD. The example shows two functions: the first one corresponds to the output of a variational circuit, $Q(z,\theta_q)$, where $z$ and $\theta_q$ are both classical inputs (e.g. classical information encoded into the circuit and variational parameters, respectively). The second function is classical (e.g. a feed-forward neural network), taking as inputs the value of the function $Q(x,\theta_q)$ and parameters $\theta_c$ and producing the output $\mathcal{L}(Q, \theta_c)$. At the computational level, these functions are implemented as programming functions or instances of a computational class and are executed separately by quantum and classical modules, respectively. The quantum module can be interpreted as a classical computer that has access to a quantum processor, while the classical module incorporates only classical computing resources. 

Suppose we want to compute $\frac{\partial \mathcal{L}}{\partial \theta_q}$ using AD. In the forward pass, $\mathcal{L}(Q, \theta_c)$ is calculated by first computing $Q(z,\theta_q)$ using the quantum module. This value is passed onto the classical module, which computes the final output given some value of $\theta_c$. In the backward pass, the classical module estimates $\frac{\partial \mathcal{L}}{\partial Q}$ and passes this information backwards to the quantum module. The quantum module estimates $\frac{\partial Q}{\partial \theta_q}$ using the techniques described in Appendix \ref{app:grad-estimation} and uses the value of $\frac{\partial \mathcal{L}}{\partial Q}$ provided by the classical module to compute $\frac{\partial \mathcal{L}}{\partial \theta_q}$ by application of the chain rule: $\frac{\partial \mathcal{L}}{\partial \theta_q} = \frac{\partial \mathcal{L}}{\partial Q} \frac{\partial Q}{\partial \theta_q}$. An analogous procedure can be applied to compute all the derivatives required for training the VQG architecture using adversarial learning (Eqs. \ref{eq:VQG-grads}-\ref{eq:VQG-grads2}). With this infrastructure in place, the optimization of all the parameters of the model can be performed using standard gradient-based optimizers such as Adam or SGD. Algorithm \ref{algorithm1} summarizes the pseudocode for the adversarial learning of the VQG model.

\section{Implementation}\label{sec:numerical-simulations}

\subsection{Numerical simulations}

\begin{figure}[H]
\centering
\includegraphics[width=15.5cm]{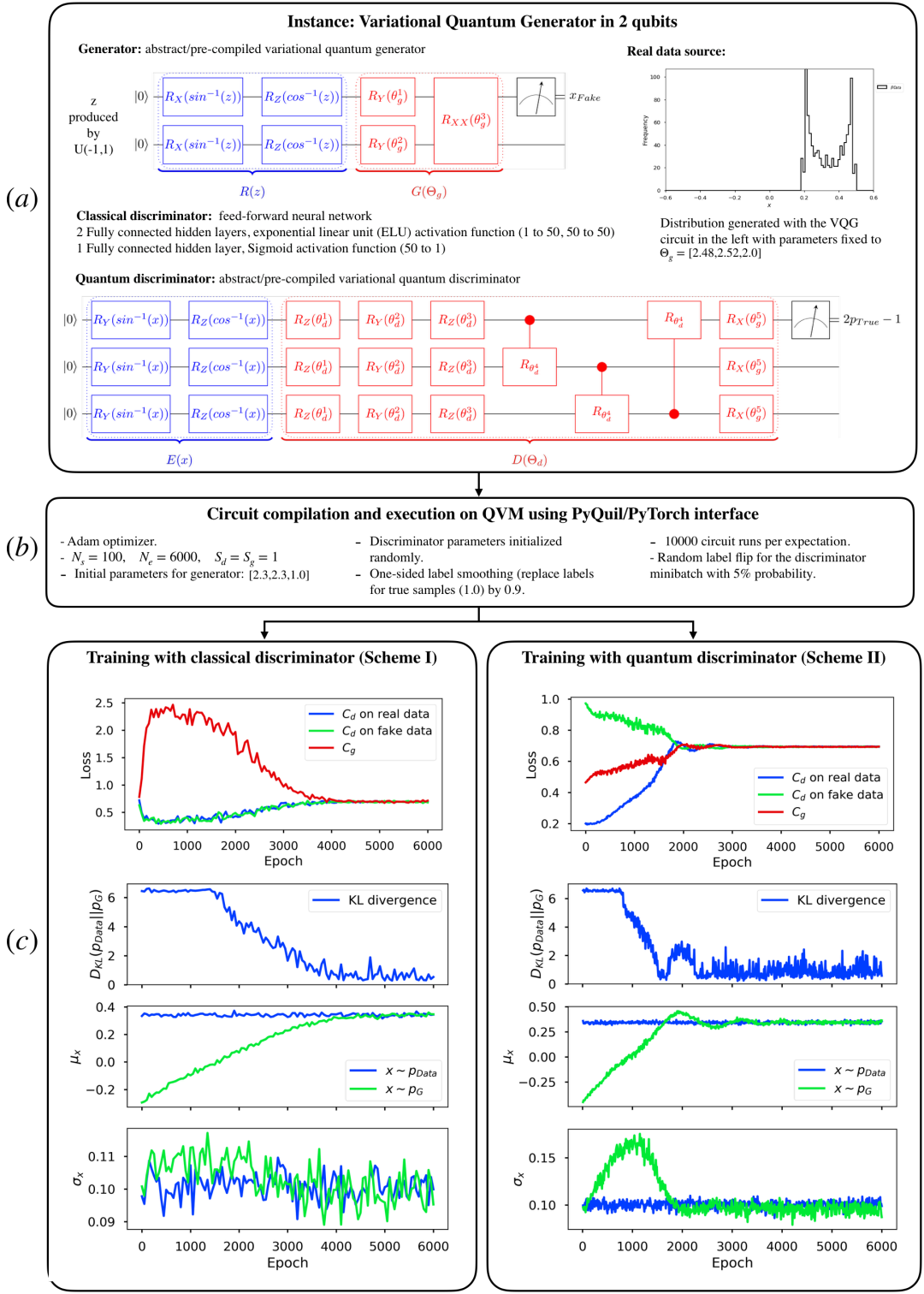}
\caption{Example of the implementation and training of a VQG instance following the \texttt{algo2qpu} approach \cite{sim2018framework}. (a) Architecture of the generator and discriminators used in the numerical experiments. The part of the circuits corresponding to encoding circuits and variational circuits are shown in blue and red, respectively. The real distribution corresponds to the quantum generator architecture with parameters fixed at $\Theta_g = [2.48, 2.52, 2.0]$ and Pauli set $[\sigma^1_X]$. (b) Details of the implementation and execution of the experiments. We performed noiseless simulations using a QVM. The generator is initialized at $\Theta_g = [2.3, 2.3, 1.0]$. (c) Training dynamics using schemes I (left panel) and II (right panel). Each panel shows from top to bottom: loss functions ($C_d$ and $C_g$) as a function of the number of epochs, Kullback-Leibler (KL) divergence between the target distribution and the generator ($D_{KL}(p_{Data}||p_{G})$), mean ($\mu_x$) and standard deviation ($\sigma_x$) of the two distributions as the optimization progresses, computed from the samples obtained at each epoch.}\label{fig:results}
\end{figure}

To illustrate the implementation of the VQG model and demonstrate its feasibility, we designed a controlled experiment where the real data source is generated by a VQG instance with the same structure as the generator. This guarantees that a solution to the learning problem exists, allowing us to focus on studying the convergence of the training process. This also facilitates the assessment of the success of the training process by directly comparing the two distributions. In our experiment, the adversarial learning process incorporates the following elements, illustrated in Figure \ref{fig:results}(a):

\begin{algorithm}[b]
\SetAlgoLined
\KwResult{Optimal $\Theta_g$, $\Omega_g$ and $\Theta_d$}
\KwData{$N_s$, $N_e$, $S_d$, $S_g$, Initial $\Theta_g$, $\Omega_g$ and $\Theta_d$\;}
 \For{$n:=1$ to $N_e$}{
 \For{$s_1:=1$ to $S_d$}{
Sample $N_s$ times from $p_z(z)$: $\{z^{(1)}, z^{(2)}, \cdots, z^{(N_s)}\}$\;
Sample $N_s$ times from the data distribution: $\{x^{(1)}, x^{(2)}, \cdots, x^{(N_s)}\}$\;
Update $\Theta_d$ by ascending discriminator's gradient: $\nabla_{\Theta_d} \frac{1}{M} \sum^{M}_{i} C_d(z^{(i)}, x^{(i)},\Theta_g, \Omega_g, \Theta_d)$ \;
 }
 \For{$s_2:=1$ to $S_g$}{
Sample $N_s$ times from $p_z(z)$: $\{z^{(1)}, z^{(2)}, \cdots, z^{(N_s)}\}$\;
Update $\Theta_g$ and $\Omega_g$ by descending generator's gradient: $\nabla_{\Theta_d} \frac{1}{N_s} \sum^{N_s}_{i} C_g(z^{(i)}, \Theta_g, \Omega_g, \Theta_d)$ \;
 }
 }\caption{\footnotesize Adversarial learning of a variational quantum generator (VQG). Training proceeds for $N_e$ epochs. At each epoch, the parameters of the discriminator and the generator are updated separately, $S_d$ and $S_g$ times respectively. Cost functions are estimated by taking $N_s$ samples of the real and the synthetic data distributions.}\label{algorithm1}
\end{algorithm}

1. \emph{Generator:} Our generator corresponds to a VQG model composed of a product encoding circuit with two layers of one qubit gates incorporating the tensorial mapping strategy to introduce non-linearities. The variational circuit of the generator is a two qubit circuit with a layer of single-qubit $Y$ rotations followed by evolution under the operator $XX$, for a total of three parameters. The measurement decoding is performed with a single operator $[\sigma^1_Z]$ without classical post-processing. This generator produces a probability distribution $p_{G}(x)$, with $x \in \mathbb{R}, \quad x \in [-1,1]$. At the beginning of the training, the variational circuit is initialized with parameters $\Theta_g = [2.3, 2.3, 1.0]$.

2. \emph{Discriminator:} We tested the two schemes described in Figure \ref{fig:HQGAN-schemes} for training the generator. In scheme I, we employed the classical feed-forward neural network described in Figure \ref{fig:results}(a) as discriminator. In scheme II, we used a quantum discriminator comprising a product encoding circuit and a variational circuit on three qubits. The variational circuit for the discriminator corresponds to a single $B(3,1)$ block encompassing a layer of arbitrary single qubit rotations, followed by parameterized C-Phase gates and finally a layer of single qubit $X$ rotations. This type of variational circuit has been used in combination with amplitude encoding for classification tasks such as MNIST \cite{Schuld2018Circuit-centricClassifiers}. As in many application of classical GANs, we chose the discriminators to be more complex than the generator, having in this case more qubits, which is equivalent to a bigger size of the hidden layer.

\onecolumngrid

\twocolumngrid

3. \emph{Real data source}: To generate the real data distribution, we employed a VQG model with the same structure as the one used in the generator and parameters fixed to $\Theta_g = [2.48, 2.52, 2.0]$. This corresponds to the classical univariate probability distribution, $p_{Data}(x)$, $x \in \mathbb{R}, \quad x \in [-1,1]$, shown in Figure \ref{fig:results}(a).

To implement adversarial learning for the VQG instances described above, we followed the \texttt{algo2qpu} framework \cite{sim2018framework}, which provides a guideline for the implementation and deployment of quantum algorithms in near-term quantum devices. We started by implementing our variational circuits using the \texttt{PyQuil} programming language \cite{smith2016practical}, part of the Forest platform which allows for deployment on both quantum virtual machines (QVM) and quantum processing units (QPUs). The functions for computing expectation values and gradients of the expectation values of variational circuits were encapsulated using the autograd function class available in the \texttt{PyTorch} library \cite{paszke2017automatic}. This enables integration with the \texttt{PyTorch} modules for implementing classical neural networks, applying automatic differentiation and performing gradient-based optimization. In our experiments, we performed simulation of the quantum circuits on the QVM and carried out adversarial learning with the Adam optimizer. We applied typical strategies employed in classical GANs to improve convergence, such as one-sided label smoothing and random flip noise for the discriminator, as described in \ref{fig:results}(b). In our numerical experiments, the expectations values produced by the generator and quantum discriminator were estimated with 10000 noiseless circuit runs per data point. The real data distribution was generated with expectation values computed up to working precision, as this plays the role of a classical data source.

\begin{figure}[t]
\centering
\includegraphics[width=8.5cm]{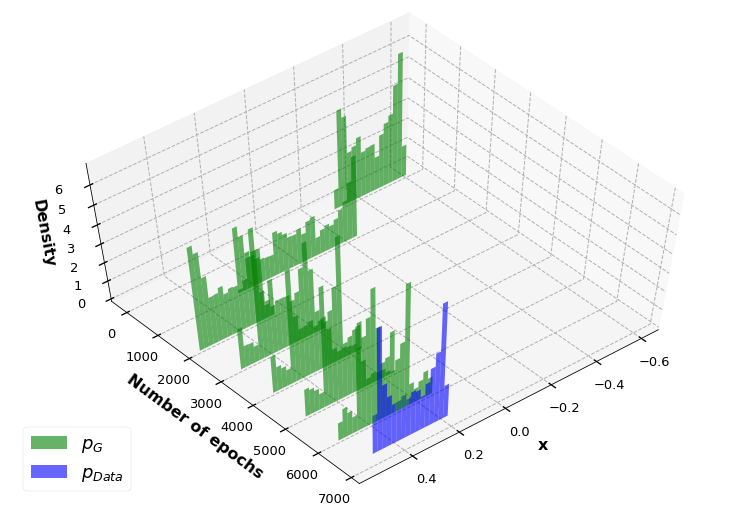}
\caption{Histogram of the data distribution produced by the generator, $p_G$, at different epochs of the training process (green histograms), compared to the target data distribution, $p_{Data}$, (blue histogram). We observe how as the optimization progresses, the generated distribution starts resembling the target one. The data corresponds to the optimization with a quantum discriminator (scheme II). All the histograms were computed using the same one thousand samples drawn from $p_z(z) \sim U(-1,1)$ as noise source.}\label{fig:histogram}
\end{figure}

Figure \ref{fig:results}(c) illustrates the training of our VQG instance with schemes I (right panel) and II (left panel). We show the dynamics of the discriminator and generator losses during training as well as the Kullback-Leibler (KL) divergence between the generated and the target distributions, computed from the discretized distributions obtained from sampling. We also track the mean and standard deviation during the optimization. For both training schemes, we observe the convergence of the losses to the expected equilibrium point located around $ln(0.5) \approx 0.7$. At the beginning of the training, the learning signal from the discriminator is relatively low and the KL divergence is mostly constant, however, it starts decreasing as the the learning signal rises. We observe that both schemes achieve convergence to an approximation of the target distributions, as evidenced by the evolution of the KL divergence and the distribution moments plotted on Figure \ref{fig:results}(c). Figure \ref{fig:histogram} compares the distribution produced by the generator with the target distribution, at different moments of the training process for scheme II.

During the simulation, we also tracked the gradients of the the discriminator and generator, noticing that the gradient components in scheme II (quantum discriminator) were around an order of magnitude larger compared to scheme I. Such large gradients can lead to convergence issues during the optimization. In particular, we observe a non-converging oscillatory behavior of the training dynamics in some of the first numerical experiments. This behavior is well documented on the classical GAN literature and is associated to the lack of an incentive for the discriminator to converge once the generator reaches the target distribution \cite{mescheder2018training}. We alleviated this problem by reducing the learning rates for the Adam algorithm, which shifted the dynamics to a damped oscillation, as the one observed in the right panel of Figure \ref{fig:results}(c). Standard approaches for treating this problem involves introducing regularization terms for the discriminator cost function on the real data \cite{mescheder2018training,roth2017stabilizing}

The observation about the magnitude of the gradients can be linked to the difference in the parameterization of variational circuits compared to neural networks. As pointed out in \cite{Schuld2018Circuit-centricClassifiers}, a variational circuits acting on $n$ qubits can be interpreted as a linear unitary layer acting on a vector of size $2^n$. Correspondingly, in the language of neural networks, this unitary can be seen as a matrix of complex weights of dimensions $2^n \times 2^n$, parameterized by only $poly(n)$ variables, corresponding to the tunable parameters of the variational circuit. In contrast, each of the $4^n$ entries of the weight matrix is a parameter in a typical layer of a fully-connected feed-forward neural network. Since gradients are calculated with respect to all the parameters, a learning signal passing through a dense layer of size $2^n \times 2^n$ is distributed among all the $4^n$ weights. In contrast, the same signal would distribute among only a polynomial number of parameters in the case of the variational circuit, leading to much larger gradient components. This comparison also offers insights into the utility of variational circuits for machine learning, as a tool for efficiently implementing linear hidden layers with high dimensions. 

\subsection{Implementation on NISQ devices}\label{subsec:nisq}

Our proposed VQG model can be implemented on a fault-tolerant quantum computer, but its variational nature makes it especially suitable for implementation on a noisy-intermediate scale quantum devices. As others HQC approaches, the cost of the algorithm is associated to the number of samples required for evaluating and training the model. As pointed out in Section \ref{subsubsec:measurement-decoding}, the repetition cost of evaluating the model scales as $O(M/\epsilon^2)$, where $M$ is the number of operators measured in the decoding step and $\epsilon$ is the precision for each expectation value. A single gradient evaluation scales as $O(n_g M/\epsilon^2)$, where $n_g$ is the number of parameterized gates in the variational circuit. For many of the variational circuits discussed here, the number of parameters is linear in the number of qubits, $n_g \in O(N)$. The minimal number of qubits required for the implementation is determined by the number of qubits required by the quantum encoder. For the product encoders used in this work, the number of qubits scales linearly with the size of the noise vector, $z$. while the depth of the encoding circuit is only constant. One could envision more general variational encoding circuits that trade circuit depth by number of qubits. In the case of amplitude encoding, the number of qubits used is only $O(log(n))$ and $O(n)$ two-qubit gates are required for state preparation.

In addition to the sampling cost, the estimation of the gradients of the variational circuits faces two important challenges: 1) the impact of noise on the estimation of expectation values and 2) the recent observation that the gradients of near-random variational circuits tend to vanish with a probability exponential on the number of qubits, an phenomenon known as \emph{barren plateau} of the quantum neural network training landscapes \cite{mcclean2018barren}. To address the issue of noise in the VQG implementation, we could apply some of the recent proposals for error mitigation in the estimation of expectation values on NISQ devices. The basic principle of these proposals is that the first order contributions of the noise to the expectation values can be removed by introducing a controllable source of noise in the circuit of interest \cite{temme2017error,endo2017practical}. The expectation values are estimated at different error levels and an extrapolation to zero noise is performed using simple regression techniques. These methods have been already applied in experiments for VQE and variational quantum classification \cite{kandala2018extending}. While error mitigation could benefit the gradient estimation, we also point out that noise is generally included in practice to improve convergence during the GAN training \cite{mescheder2018training}.
Most likely, the training process will be able to tolerate moderate levels of noise, as observed in the case of variational circuits and tensor networks employed in classifications tasks \cite{huggins2018towards,grant2018hierarchical,Schuld2018Circuit-centricClassifiers}. Error mitigation will likely play a more crucial to generate high quality samples after training is complete.

In our proof of principle experiments, we did not observe vanishing gradients most likely due to the small size of the circuits used in the example. In larger scale implementation of the VQG model, barren plateaus might become an issue. In this case, several strategies could be employed to mitigate the problem. One of them, especially suitable for variational circuits built on circuit blocks, is \emph{block-by-block} training. In this case, the optimization starts with a variational circuit with a single or a few circuit blocks, which are less likely to suffer from the barren plateau issue due to the relatively small number of parameters. In subsequent rounds, we add more blocks to the variational circuit and use the optimal variational parameters of the previous round to initialize the new round of training. This procedure can improve convergence, as shown in the case of classical deep neural networks \cite{bengio2007greedy,he2016deep}. 

Another strategy is to use circuits with subcomponents that admit classical simulation or inspired by classically simulable circuits. An example of such circuits is the low-depth circuit ansatz (LDCA) proposed in Ref.\cite{dallaire2018low} for quantum simulation of fermions. The basic building block of the circuit is composed of matchgates \cite{valiant2002quantum}, that can be simulated classically, augmented with a set $ZZ$ rotations that increase the complexity of the circuit. Therefore, we could run classical simulations of the VQG training with an LDCA variational circuits without the $ZZ$ interactions. The optimal parameters obtained from the classical simulation can be then employed to initialize the training with the full LDCA circuit using the quantum computer. A similar procedure can be applied to variational circuits based on tensor networks, that admit simulation on classical computers with small bond dimensions, as suggested in Refs. \cite{huggins2018towards,grant2018hierarchical}. 

\section{Conclusion and outlook}\label{sec:conclusions}

In this paper we have presented a hybrid quantum-classical architecture, comprising variational quantum circuits and classical functions, for modeling continuous probability distributions. Our variational quantum generator incorporates two quantum circuits: a quantum circuit encoding a classical random variable into a quantum state, $R(z)$, and a variational circuit $G(\Theta_g)$, whose parameters are optimized to mimic the target classical probability distribution. A sample, $x_{Fake}$ from the VQG architecture is generated sampling $z$ from a noise distribution $z\sim p_z(z)$, encoding this variable into a quantum state using the encoding circuit, applying the variational circuit and measuring the expectation values of a set of predefined operators. The vector of expectation values, $P$, can be post-processed using a classical function, $f_g(P)$, such as a neural network, to generate the sample $f_g(P)=x_{Fake}$. The VQG architecture can be trained using a gradient-based adversarial learning strategy, where a second model, known as discriminator, compares the quality of the samples generated by the VQG model with samples from the real data distribution. We show that the required gradients can be calculated using existing techniques for evaluating gradients of variational functions and exploiting the established infrastructure for automatic differentiation of classical functions. We illustrate this process with a simple proof-of-principle experiment where a VQG instance with fixed parameters serves as the target distribution.

Our proposal contributes to an increasing body of work exploring the use of hybrid quantum-classical computing in machine learning, offering an approach to perform generative modeling of continuous probability distributions with quantum computers. Furthermore, the same architecture employed in VQG can be used to build models for classification, as illustrated in Section \ref{subsec:costF}. We also present a strategy for training our proposed generator with both classical and quantum discriminators, taking advantage of the integration of gradient estimation of variational circuits and automatic differentiation strategies. The incorporation of these tools can also benefit the implementation of other HQC algorithms, such as VQE. Nowadays, the availability of libraries for programming and executing quantum circuits \cite{smith2016practical,Cirq,Qiskit,ProjectQ} that are compatible with standard libraries for machine learning \cite{paszke2017automatic,tensorflow2015-whitepaper}, facilitate this integration. Recently, specialized libraries for automatic differentiation of variational circuits has been also developed \cite{bergholm2018pennylane}.

There are several open questions that remain to be investigated. Perhaps the most significant one is whether this approach can offer an advantage with respect to purely classical models for generative learning. Some theoretical works \cite{Boixo2016CharacterizingDevices,du2018expressive} indicate that variational circuits might bear an advantage for discrete generative tasks, however the extent to which this impacts practical applications such as image, sound and language generation will require extensive computational studies on real instances. A second aspect that needs to be studied is the role of noise the performance of the VQG model implemented on NISQ devices. As discussed in Section \ref{subsec:nisq}, one of the strategies to improve convergence during classical GAN training is to introduce noise in the data, which generally prevents over-fitting in the discriminator and improve the robustness of the final model. The extent to which noise on NISQ devices can be tolerated or can benefit the training process, as well as the overall quality of the the distributions generated by VQG, most likely depends on the specific noise process and the nature of the target distribution. Finally, a third research direction is the study of adversarial learning of classical generators using quantum discriminators, in particular how this particular arrangement could impact the convergence of the training process.

An important advantage of the VQG model is its flexibility, allowing for exploring multiple choices of variational circuits and encodings. This also allows for designing new strategies to incorporate non-linearities through hybrid quantum-classical architectures. The architecture presented in this work incorporates non-linearities through pre-processing classical data with non-linear functions and through the tensorial mapping approach. The variational circuit acts as a linear layer but non-linearities can be incorporated after measurement via classical post-processing. Both, variational circuits and classical post-processing can be considered together as a single non-linear layer and can be repeated to build deep hybrid quantum-classical architectures. Another important future research direction is the design of new circuits for variational encodings that can balance the cost of number of qubits and depth while incorporating non-linearities. In the specific case of VQG, the quantum encoder, $R(z)$, plays the role of noise source, sampling randomly a state from the manifold defined by the encoding circuit. An alternative strategy could be replacing the encoding circuit with an efficient circuit to approximately sample from the Haar measure, for example, an efficiently implementable unitary 2-design \cite{dankert2009exact}.

Finally, the most thrilling aspect of the VQG approach is the prospect of realizations of the algorithm on quantum devices to tackle standard problems in generative learning. As NISQ devices approach sizes that surpass the possibility of classical simulation, quantum algorithms that allow for gradually incorporating quantum capabilities into the established quantum machine learning pipelines will be required. The VQG model offers such a framework, establishing a strategy to combine increasingly large variational circuits with standard neural networks to model data distributions. Future work will be dedicated to exploring the utility of the VQG approach in specific scientific and industrial applications, including image processing, finance, medicine, cybersecurity, and drug and materials design.

\section{Acknowledgements}
We thank Sukin Sim and Max Radin for their assistance in the implementation of the numerical experiments. We also thank Sukin Sim, Yudong Cao, Peter Johnson, Pierre-Luc Dellaire-Demers and Max Radin for helpful discussions and suggestions to the manuscript. The authors acknowledge support from the Army Research Office under Award No.\ W911NF-15-1-0256 and Office of Naval Research under award N00014-16-1-2008. AA-G also acknowledges support from the Vannevar Bush Faculty Fellowship program sponsored by the Basic Research Office of the Assistant Secretary of Defense for Research and Engineering, under Award No.\ ONR 00014-16-1-2008, and generous support from Anders G. Fr\o seth and the Canada 150 Research Chair Program.

\appendix

\section{Variational circuits architectures}\label{app:VCs}

\begin{figure}[t]
\begin{center}
\includegraphics[width=8cm]{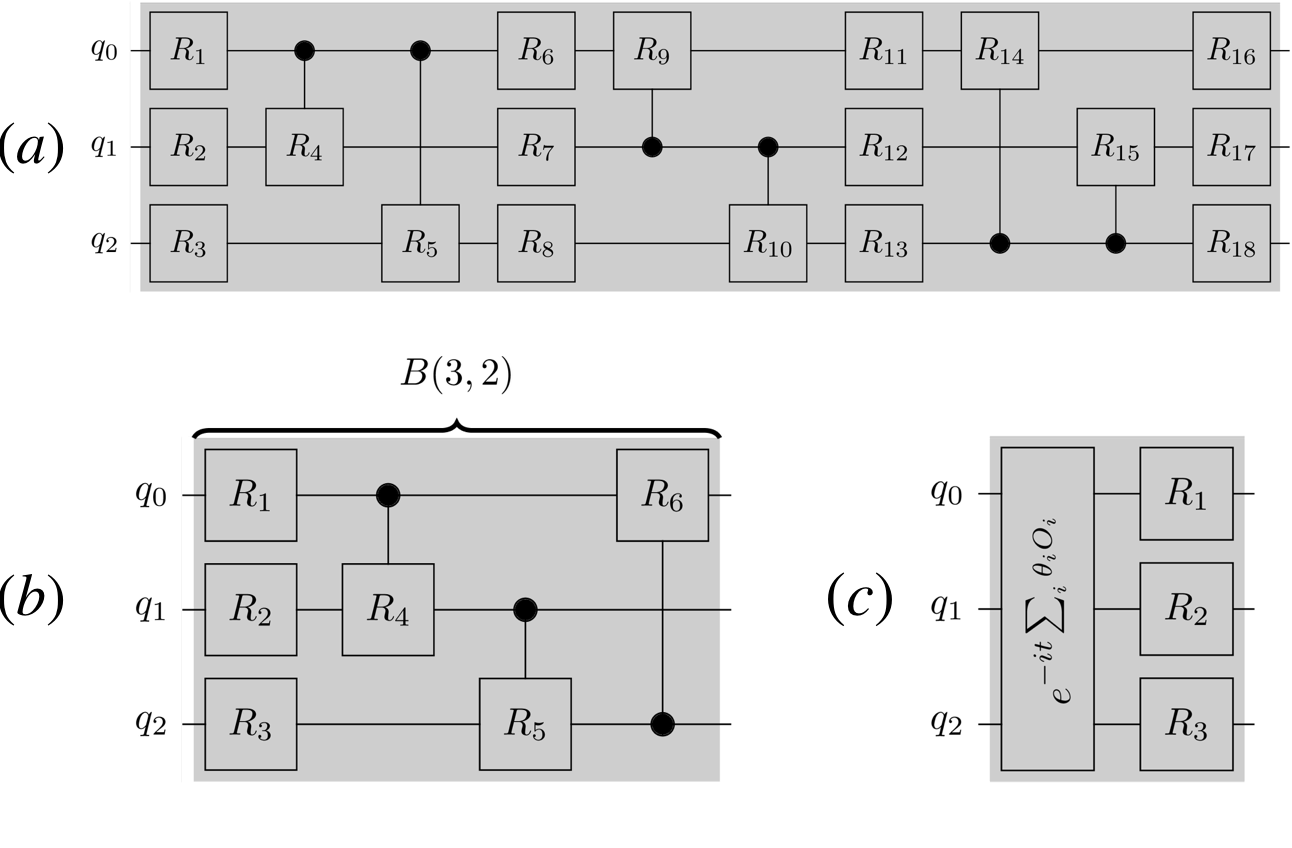}
\end{center}
\caption{Examples of variational circuit blocks employed in quantum machine learning: a) Circuit block employed for HQC algorithms such as QAE and QVECTOR \cite{Romero2016QuantumData,johnson2017qvector}. b) Generalization of circuit blocks proposed by Schuld et al. \cite{Schuld2018Circuit-centricClassifiers}. The depicted blocks corresponds to 3 qubits with range 2 ($B(3,2)$). c) Variational circuit corresponding to quantum evolution under a Hamiltonian with tunable parameters ($\theta_j$) and single-qubit rotations. $R_j$ represents a generic single-qubit gate.}
\label{discriminativeVC}
\end{figure}

Figure \ref{discriminativeVC} describes some examples of variational circuits employed in HQC algorithms. Figure \ref{discriminativeVC}(a) shows a circuit block containing all the possible controlled one-qubit rotations among a set of qubits, interleaved with a set of single qubit rotations. We start considering the rotations controlled by the first qubit, followed by the rotations controlled by the second qubit and so on. The number of parameters in this circuit block scales as $O(n^2)$. More simplified circuit blocks, where entangling operations are not parameterized, has been used for experimental demonstration of QML for classification \cite{Havlicek2018SupervisedSpaces}. In this case, the disposition of the entangling gates is generally dictated by the constraints in the qubit connectivity of the processor.

Families of circuit blocks have been also proposed. In particular, Schuld et al. proposed a series of circuit blocks for classification, generically referred to as code blocks $B(n,r)$g \cite{Schuld2018Circuit-centricClassifiers}. An example of these blocks is depicted in \ref{discriminativeVC}(b). A block $B(n,r)$ comprises a layer of general single-qubit rotations $R=R(\alpha,\beta,\gamma)$ applied to each of the $n$ qubits of the register followed by a layer of $n/\text{gcd}(n,r)$ controlled-$R$ gates, where $r$ is the range of the two-qubit gates and gcd$(n,r)$ is the greatest common divisor of $n$ and $r$. The target and control qubits of the j-th two-qubit gate in the block are given by the integers $t_j = (jr-r)\mod n$ and $c_j = jr \mod n$, respectively. This definition guarantees a number of parameters that scales linearly with the size of the qubit register, $n$. In general, B-blocks are capable of entangling/unentangling all the qubits with numbers that are multiple of gcd$(n,r)$. If $n$ and $r$ are co-prime, the network of entangling gates forms a cycle graph capable of entangling/unentangling all the qubits in the register. Finally, \ref{discriminativeVC}(c) presents and example of a generic circuit implementing evolution under a local Hamiltonian, where the coefficient of the Hamiltonian terms serve as variational parameters.

\section{Estimation of analytical gradients for variational circuits}\label{app:grad-estimation}

\begin{figure*}[t]
\begin{tabular}{c}
\Qcircuit @C=0.7em @R=1.0em {
|0\rangle  && \qw & \qw & \qw & \gate{H} & \qw & \ctrl{1} & \qw & \qw & \qw & \qw & \qw & \ctrl{1} & \gate{H} & \gate{X(-\frac{\pi}{2})}&\meter \\
|0\rangle^{\otimes n} && \qw {/} & \gate{U_1(\theta^1)} & \qw & \cdots && \gate{V_j} & \gate{U_{j}(\theta^{j})} &\qw  & \cdots && \gate{U_{N_g}(\theta^{N_g})} & \gate{P_i} & \qw & \qw \\ 
}
\end{tabular}
\caption{Circuit for measuring $\frac{\partial \langle 0| U(\Theta)^{\dagger} P_i  U(\Theta) |0\rangle}{\partial \theta^j}$. The gate $X(\frac{\pi}{2})$ rotates to the $Y$-basis.}\label{fig:gradcirc}
\end{figure*}
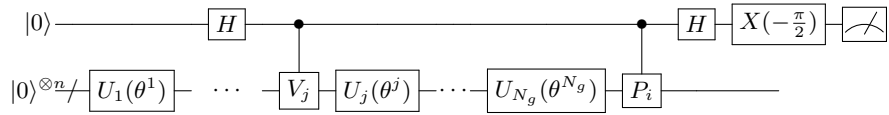

Recently, it has been shown that gradients of expectation values of variational circuits can be estimated analytically using slight modifications of the initial quantum circuit. Specifically, consider a variational circuit $U(\Theta)$ of the form:
\begin{equation}
U(\Theta) = U_1(\theta^1)U_2(\theta^2)\cdots U_{N_g}(\theta^{|\Theta|}),
\end{equation}
where $U_j(\theta^j) = \exp(-i \theta^j V_j / 2)$ with $\{ V_j \}$ being a Pauli operator. One strategy for computing gradients with respect to the parameter $\theta^i$ is using the circuit of Figure \ref{fig:gradcirc}, which requires one additional qubit compared to the original variational circuit \cite{Romero2017StrategiesAnsatz,Schuld2018Circuit-centricClassifiers}. Taking measurements on this ancilla qubit provides an estimate to one element of the Jacobian of the vector $P$, $\nabla P_{i,j}$. 

Alternatively, the same component can be estimated using two separate evaluations of the expectation values where the original variational circuit is replaced by the modified circuits $^{+}U_j(\Theta)$ and $^{-}U_j(\Theta)$ \cite{Mitarai2018QuantumLearning} defined as:
\begin{align}
&^{+}U_j(\Theta) = U(\theta^1)\cdots U_j \left( \theta^j + \frac{\pi}{2} \right) \cdots U(\theta^{|\Theta|}), \\
&^{-}U_j(\Theta) = U(\theta^1)\cdots U_j \left( \theta^j -\frac{\pi}{2} \right) \cdots U(\theta^{|\Theta|}).
\end{align}
such as:
\begin{align}
&\frac{\partial \langle U(\Theta)^{\dagger} |P_i|  U(\Theta) \rangle}{\partial \theta^j} = \notag \\& \text{Re} \left( \langle ^{+}U_j(\Theta)^{\dagger} |P_i|  ^{+}U_j(\Theta) \rangle -
\langle ^{-}U_j(\Theta)^{\dagger} |P_i|  ^{-}U_j(\Theta) \rangle \right)
\end{align}
This approach has been recently coined as a classical linear combination of unitaries (CLCU) \cite{schuld2018evaluating}. Compared to the circuit of Figure \ref{fig:gradcirc}, the CLCU strategy requires twice as many measurements to estimate the gradient to the same accuracy. However, it does not require an ancilla qubit and employs the same circuit as the objective function, $U(\Theta)$. The later implies that the compiled of $^{+}U(\Theta)$ and $^{-}U(\Theta)$ are the same as $U(\Theta)$, which might simplify the implementation.

\section{Automatic differentiation}\label{app:AD}

Automatic differentiation (AD) \cite{neidinger2010introduction,du2018expressive} is an algorithmic strategy to extend a program that computes numerical values of a function such as it can also compute arbitrary derivatives of the same function. Unlike numerical differentiation, AD provides exact derivatives up to working computational precision. AD also differs from symbolic differentiation in the sense that it computes numerical values of the derivatives instead of analytical expressions. To achieve its goal, AD extends the domain of variables in the computation to incorporate derivative values and introduces a programming semantics to enable the propagation of the derivatives using the chain rule. This allows to compute arbitrary derivatives of any function by applying differentiation to the sequence of elementary operations and elementary functions that implement the function on the computer. The process is performed automatically during execution time and has only a constant overhead in computational cost compared to the execution of the original function.

There exist different strategies for implementing AD. For this paper, we implemented a strategy known as reverse mode accumulation \cite{du2018expressive}, which is a generalization of the back-propagation procedure employed in feed-forward neural networks \cite{goodfellow2016deep}. In this case, the calculation of a function is broken down into a series of intermediate steps with results stored by intermediate variables. The inter-dependence of these variables constitutes a directed graph known as computational graph. In reverse mode, the function and its derivatives are calculated in two steps known as the \emph{forward} and \emph{backward} passes. In the forward pass, the original function code is run, computing the values of all the intermediate variables and recording the dependencies of the computational graph using a book-keeping procedure. In the backward pass, derivatives are calculated by computing the derivatives of each intermediate variable with respect to its immediate inputs and propagating the derivatives in reverse through the computational graph, from outputs to inputs. 

\bibliographystyle{apsrev}
\bibliography{ref.bib}

\end{document}